# The Precise Inner Solutions of Gravity field Equations of Hollow and Solid Spheres and the Theorem of Singularity

Mei Xiaochun

( Institute of Innovative Physics in Fuzhou, Department of Physics, Fuzhou University)

**Abstract** In the present calculation of the inner solution of gravity field equation with spherical symmetry based on general relativity, in order to avoid the singularity appearing in the center of sphere, we actually let the integral constant to be zero. It is proved in this paper that the constant can not be zero. According to the theory of differential equation, the integral constant should be determined by the boundary conditions of spherical surfaces, in stead of the metric in the spherical center. Meanwhile, the spherical volume of three dimensional in curved space is different from that in flat space. By considering these two factors, the integral constant can not be zero. The metric of inner gravity field of hollow sphere is calculated at first. Then let the inner radius of hollow sphere become zero, we obtain the metric of inner gravity field of solid sphere. Based on the definition of energy momentum tensor of general relativity, the gravity masses of hollow and solid spheres in curved space are calculated strictly. The relations between gravity masses and spherical radius are obtained. The results indicate that no matter what the masses and densities of hollow sphere and solid sphere are, space-time singularities would appear in the centers of spheres. Meanwhile, no matter what the mass and density are, the intensity of pressure at the center point of solid sphere can not be infinite. That is to say, the material can not collapse towards the center of so-called black hole. In stead, it may be that there exist the spherical surfaces of infinite pressure inside the hollow and solid spheres, and material would collapse toward the surfaces so that the common spheres are unsteady. At the center of solid sphere and on its neighboring region, pressure intensities would become negative values. There may be a region for hollow sphere in which pressure intensities would become negative values too. On the inner and external surfaces of hollow sphere, pressure intensities would not be zero. The common hollow and solid spheres in daily live would have such impenetrable characteristics. The results only indicate that the singularity black holes predicated by general relativity are caused by the descriptive method of curved space-time and can not exist in nature actually. If black holes exist really in the universe, they can only be the Newtonian black holes, not the Einstein's black holes. The relation between the conclusions of this paper and the Hawking theorem of singularity is discussed at last. It is pointed out that the results revealed in the paper are consistent with the theorem of singularity. Three preconditions of the theorem of singularity do not contain the restriction on the masses and densities of material systems. According to general relativity, it is an inevitable result for the singularities to appear in the centers of hollow and solid spheres composed of common material with small densities and masses.

Black hole, Theorem of singularity

## 1．The Schwarzschild inner solution of gravity field equation

We know that the strict solutions of the Einstein's equation of gravity field with spherical symmetry are the Schwarzschild solutions including inner and external solutions. We consider a static and uniform sphere with radius $r_0$ and constant density $\rho_0$, inner pressure intensity $p(r)$ is related to coordinate but does not depends on time. By considering the static energy momentum tensor of idea fluid, the Schwarzschild inner solutions is [1]

$$ds^2 = c^2 \left( \frac{3}{2}\sqrt{1-\frac{r_0^2}{R^2}} - \frac{1}{2}\sqrt{1-\frac{r^2}{R^2}} \right)^2 dt^2 - \frac{dr^2}{1-r^2/R^2} - r^2(d\theta^2 + \sin^2\theta d\varphi^2) \qquad (1)$$

Here $R^2 = 3c^2/8\pi G\rho_0$. The metric is finite at the center point of sphere. However, it should be pointed out that in the process of solving the Einstein's equation of gravity field what we obtain is actually

$$g_{11}(r) = \left(1 - \frac{r^2}{R^2} + \frac{A}{r}\right)^{-1} \qquad (2)$$

The metric has a singularity at point $r = 0$. In order to eliminate the singularity, we let integral constant $A = 0$ directly in the current theory. It is proved in this paper that this is irrational. According to the theory of differential equation, integral constant should be determined by the boundary conditions of spherical surfaces, in stead of the metric in the spherical center. Because the volume of sphere in curved space is different from that in flat space, by considering the continuity conditions on the surfaces of spheres, we inevitably reach the result with $A \neq 0$. Therefore, no matter what the masses and density of solid sphere are, the singularities at the center of sphere are inevitable. On the other hand, according to the current theory, the inner pressure intensity of sphere is [2]

$$p(r) = \rho_0 c^2 \frac{\sqrt{1-(r/R)^2} - \sqrt{1-(r_0/R)^2}}{3\sqrt{1-(r_0/R)^2} - \sqrt{1-(r/R)^2}} \qquad (3)$$

On the spherical surface $r = r_0$ we have $p(r_0) = 0$. In order to make pressure intensity finite at the center of sphere with $r = 0$, we have to introduce a constraint condition for spherical radius

$$r_0^2 < \frac{8}{9}R^2 \qquad \text{or} \qquad r_0 > \frac{9}{8}r_g = \frac{9GM}{4c^2} \qquad (4)$$

Here $r_g = 2GM/c^2$ is the Schwarzschild radius. If $r_0 \leq 9r_g/8$, pressure intensity becomes infinite. In this way, stable solution becomes impossible and material would collapse towards the center of sphere so that singularity black holes appear.

However, if integral constant $A \neq 0$, pressure intensity (3) and constraint condition (4) will be changed. All calculations about high density celestial bodies including black holes based on (3) and (4) in the current astrophysics should be reconsidered.

Let's first calculate the solutions of gravity field equations of hollow and solid spheres strictly, secondly discuss the singularity problems of hollow and solid spheres below.

## 2. The strict solution of inner gravity field of hollow sphere

Suppose that the inner radius of hollow sphere is $R_1$ and external radius is $R_2$, the gravity mass is $M$. How to define gravity mass is very important in this paper, we will discuss this problem in detail later. The region $I_3$ with $r > R_2$ and the region $I_1$ with $r < R_1$ are vacuum. The region $I_2$ beneath two spherical shells with $R_1 < r < R_2$ is composed of complete liquid with constant density $\rho_0$ and pressure intensity $p(r)$. Because the material is distributed with spherical symmetry, the space-time metric can be written as

$$ds^2 = c^2 e^{\nu(r)} dt^2 - e^{\lambda(r)} dr^2 - r^2 \left( d\theta^2 + \sin^2\theta \, d\varphi^2 \right) \tag{5}$$

In the vacuum region $I_3$, the solution of the Einstein's equation of gravity field is the well-known Schwarzschild solution. We can write it as

$$g_{00}(r) = e^{\nu(r)} = \left(1 + \frac{A_3}{r}\right) \qquad g_{11}(r) = -e^{\lambda(r)} = -\left(1 + \frac{A_3}{r}\right)^{-1} \tag{6}$$

I order to determine integral constant $A$, we have to compare it with the Newtonian theory under the asymptotic condition with $r \to \infty$

$$g_{00}(r)\big|_{r \to \infty} = 1 + \frac{2\psi(r)}{c^2} \qquad \psi(r) = -\frac{GM}{r} \tag{7}$$

Here $M$ is the static gravity mass of the hollow sphere in the Newtonian theory. By comparing (6) with (7), we obtain $A_3 = -2GM/c^2$. So in the region $I_3$, we have the same result for hollow sphere

$$g_{00}(r) = \left(1 - \frac{2GM}{c^2 r}\right) \qquad g_{11}(r) = -\left(1 - \frac{2GM}{c^2 r}\right)^{-1} \tag{8}$$

To obtain the metric of the region $I_2$ beneath two spherical shells, the mixing energy momentum tensor of complete fluid is used [1]:

$$T_0^0 = \rho_0 c^2 \qquad T_1^1 = T_2^2 = T_3^3 = -p_0 \qquad R_1 \leq r \leq R_2 \tag{9}$$

We substitute (5) and (9) in the Einstein's equation of gravity field

$$R_\mu^\nu - \frac{1}{2} g_\mu^\nu R = -\frac{8\pi G}{c^4} T_\mu^\nu \tag{10}$$

According to the standard procedure of calculation in general relativity, we obtain

$$e^{-\lambda(r)} \left( \frac{\lambda'}{r} - \frac{1}{r^2} \right) + \frac{1}{r^2} = \frac{8\pi G \rho_0 c^2}{c^4} \tag{11}$$

$$e^{-\lambda(r)} \left( \frac{\nu'}{r} + \frac{1}{r^2} \right) - \frac{1}{r^2} = \frac{8\pi G p(r)}{c^4} \tag{12}$$

$$e^{-\lambda(r)} \left( \frac{\nu''}{2} - \frac{\lambda' \nu'}{4} + \frac{\nu'^2}{4} + \frac{\nu' - \lambda'}{2r} \right) = \frac{8\pi G p(r)}{c^4} \tag{13}$$

By calculating the integral of (11) and let $R^2 = 3c^2 / 8\pi G \rho_0$, we get

$$g_{11}(r) = -e^{\lambda(r)} = -\left(1 - \frac{r^2}{R^2} + \frac{A_2}{r}\right)^{-1} \tag{14}$$

In which $A_2$ is an integral constant. We will prove $A_2 \neq 0$ in the next section. By considering (12) minus (13), then multiplied $2/r$ on the result and considering (11), we obtain

$$\frac{dp(r)}{dr} + \left[\rho_0 c^2 + p(r)\right] \frac{\nu'(r)}{2} = 0 \tag{15}$$

The integral of (15) is

$$\rho_0 c^2 + p(r) = B_2 e^{-v(r)/2} \tag{16}$$

Here $B_2$ is an integral constant. Substituting $\rho_0$ in (11) and $p(r)$ in (12) into (16), we obtain

$$e^{v(r)/2} e^{-\lambda(r)} \left[ \frac{\lambda'(r) + v'(r)}{r} \right] = \frac{c^4 B_2}{4\pi G} \tag{17}$$

On the other hand, by the differential of (14), we have

$$\lambda'(r) e^{-\lambda(r)} = \frac{2r}{R^2} + \frac{A_2}{r^2} \tag{18}$$

Substituting (14) and (18) in (17) we get

$$e^{v(r)/2} \left[ \frac{2}{R^2} + \frac{A_2}{r^3} + v'(r) \left( \frac{1}{r} - \frac{r}{R^2} + \frac{A_2}{r^2} \right) \right] = \frac{c^4 B_2}{4\pi G} \tag{19}$$

By considering relation $e^{v(r)/2} v'(r) = 2 d e^{v(r)/2} / dr$, (19) can be written as

$$\frac{d e^{v(r)/2}}{dr} + P(r) e^{v(r)/2} = B_2 K(r) \tag{20}$$

Here $\qquad P(r) = \frac{1}{2r} - \frac{1 - 3r^3/R^2}{2(A_2 + r - r^3/R^2)} \qquad K(r) = \frac{c^4 r^2}{8\pi G(A_2 + r - r^3/R^2)} \tag{21}$

The integral of (20) is

$$e^{v(r)} = \left[ e^{-\int P(r)dr} \left( B_2 \int e^{\int P(r)dr} K(r) dr + C_2 \right) \right]^2 \tag{22}$$

Here $C_2$ is an integral constant. If let $A_2 = 0$ in (14) and (21), we reach the result in the current general relativity

$$g_{00}(r) = e^{v(r)} = \left( B - C \sqrt{1 - \frac{r^2}{R^2}} \right)^2 \qquad g_{11}(r) = -e^{\lambda(r)} = -\left( 1 - \frac{r^2}{R^2} \right)^{-1} \tag{23}$$

In which constants $B = c^4 B_2 R^2 / (8\pi G)$ and $C = -C_2$. If $A_2 \neq 0$, we have

$$e^{-\int P(r)dr} = e^{-\frac{\ln r}{2}} e^{\frac{1}{2}\int Q(r)dr} = \frac{1}{\sqrt{r}} e^{-\frac{1}{2}\int Q(r)dr} \tag{24}$$

Here $\qquad Q(r) = \frac{1 - 3r^3/R^2}{A_2 + r - r^3/R^2} = 3 + \frac{1 - 3A_2 - 3r}{A_2 + r - r^3/R^2} \tag{25}$

Let $\qquad \dfrac{1}{A_2 + r - r^3/R^2} = \dfrac{\alpha_0}{(\alpha_1 + \alpha_2 r)} + \dfrac{\alpha_3 + \alpha_4 r}{(\alpha_5 + \alpha_6 r + \alpha_7 r^2)} \tag{26}$

In which constants $\alpha_i = \alpha_i(A_2, R) \neq 0$. The forms of $\alpha_i$ are complex, but it is unnecessary to write out. We have

$$\int Q(r) dr = 3 \left( 1 - \frac{\alpha_0}{\alpha_2} \right) r + \left[ \frac{\alpha_0 (1 - 3A_2)}{\alpha_2} + \frac{3\alpha_0 \alpha_1}{\alpha_2^2} \right] \ln(\alpha_1 + \alpha_2 r)$$

$$-\frac{3}{2\alpha_7}\ln(\alpha_5+\alpha_6 r+\alpha_7 r^2)+\frac{1-3A_2+\alpha_6/(2\alpha_7)}{\sqrt{\alpha_6^2-4\alpha_5\alpha_7}}\ln\frac{2\alpha_7 r+\alpha_6-\sqrt{\alpha_6^2-4\alpha_5\alpha_7}}{2\alpha_7 r+\alpha_6+\sqrt{\alpha_6^2-4\alpha_5\alpha_7}} \quad (27)$$

Let
$$F(r)=e^{\frac{1}{2}\int Q(r)dr}=e^{3(1-\alpha_0/\alpha_2)r/2}(\alpha_1+\alpha_2 r)^{[\alpha_0(1-3A_2)/\alpha_2+3\alpha_0\alpha_1/\alpha_2^2]}$$

$$\times(\alpha_5+\alpha_6 r+\alpha_7 r^2)^{-3/(2\alpha_7)}\left(\ln\frac{2\alpha_7 r+\alpha_6-\sqrt{\alpha_6^2-4\alpha_5\alpha_7}}{2\alpha_7 r+\alpha_6+\sqrt{\alpha_6^2-4\alpha_5\alpha_7}}\right)^{\frac{1-3A_2+\alpha_6/(2\alpha_7)}{\sqrt{\alpha_6^2-4\alpha_5\alpha_7}}} \quad (28)$$

As well as
$$D(r)=\int\frac{\sqrt{r}K(r)}{F(r)}dr \quad (29)$$

Let $r=0$ in (28), we get

$$F(0)=(\alpha_1)^{[\alpha_0(1-3A_2)/\alpha_2+3\alpha_0\alpha_1/\alpha_2^2]}(\alpha_5)^{-3/(2\alpha_7)}\left(\ln\frac{\alpha_6-\sqrt{\alpha_6^2-4\alpha_5\alpha_7}}{\alpha_6+\sqrt{\alpha_6^2-4\alpha_5\alpha_7}}\right)^{\frac{1-3A_2+\alpha_6/(2\alpha_7)}{\sqrt{\alpha_6^2-4\alpha_5\alpha_7}}}\neq 0 \quad (30)$$

Therefore, when $r=0$, $D(r)$ is finite. We can write (22) as

$$e^{\nu(r)}=\frac{F^2(r)}{r}\left[C_2+B_2 D(r)\right]^2 \quad (31)$$

In the region $I_2$ beneath two spherical shells, the metric can be written as

$$ds^2=\frac{c^2 F^2(r)}{r}\left[C_2+B_2 D(r)\right]^2 dt^2-\left(1-\frac{r^2}{R^2}+\frac{A_2}{r}\right)^{-1}dr^2-r^2(d\theta^2+\sin^2\theta\, d\varphi^2) \quad (32)$$

In the vacuum region $I_1$ of hollow sphere cavity, the solution of the Einstein's equation of gravity field is still the Schwarzschild solution

$$g_{00}(r)=e^{\nu(r)}=\left(1+\frac{A_1}{r}\right) \qquad g_{11}(r)=-e^{\lambda(r)}=-\left(1+\frac{A_1}{r}\right)^{-1} \quad (33)$$

Now we obtain all solutions. Let's determine the integral constants $A_1$, $A_2$, $B_2$ and $C_2$ below.

## 3. The calculations of integral constants and gravity mass

By considering the continuity of metric tensors on the external spherical surface $r=R_2$ for $g_{11}(r)$, according to (8) and (14), we have

$$1-\frac{2GM}{c^2 R_2}=1-\frac{R_2^2}{R^2}+\frac{A_2}{R_2} \quad (34)$$

$$A_2=-\frac{2GM}{c^2}+\frac{R_2^3}{R^2}=-\frac{2GM}{c^2}+\frac{8\pi G\rho_0 R_2^3}{3c^2} \quad (35)$$

For hollow sphere, it is obvious that $A_2\neq 0$. On the external spherical surface with $r=R_2$ for $g_{00}(r)$, according to (8) and (31), we have

$$1 - \frac{2GM}{c^2 R_2} = \frac{F^2(R_2)}{R_2}\left[C_2 + B_2 D(R_2)\right]^2 \tag{36}$$

Similarly, on the internal sphere surface with $r = R_1$, according to (14), (33), (35) and (36), we have

$$1 + \frac{A_1}{R_1} = \frac{F^2(R_1)}{R_1}\left[C_2 + B_2 D(R_1)\right]^2 \tag{37}$$

$$1 + \frac{A_1}{R_1} = 1 - \frac{R_1^2}{R^2} - \frac{2GM/c^2 - R_2^3/R^2}{R_1} \tag{38}$$

From (38), we obtain

$$A_1 = -\frac{2GM}{c^2} + \frac{R_2^3 - R_1^3}{R^2} = -\frac{2GM}{c^2} + \frac{8\pi G \rho_0}{3c^2}\left(R_2^3 - R_1^3\right) \tag{39}$$

Let's now prove $A_1 \neq 0$. If space is curved, the relation between mass and volume of hollow sphere is

$$V_0 = \frac{4\pi G}{3}\left(R_2^3 - R_1^3\right) = V_0 \qquad M = \rho_0 V_0 \tag{40}$$

Here $V_0$ is the volume of hollow sphere. It should be emphasized that $M$ is the Newtonian gravity mass. We introduce it by considering the asymptotic relation (7) between the Einstein's theory and the Newtonian theory of gravity. Substitute (40) in (39), we get $A_1 = 0$. This is just the result in the current general relativity. However, (40) can not hold in curved space. Because there is length contraction along the direction of radius, we have $dl' = dl/\sqrt{-g_{11}(r)}$ and $d\sigma' = d\sigma$, so $dV' = dl'd\sigma' = dV/\sqrt{-g_{11}(r)}$. In the curved space, the volume should be calculated by the following formula [1]

$$V' = \int dV' = \int \frac{dV}{\sqrt{-g_{11}(r)}} = \int_{R_1}^{R_2} \frac{4\pi r^2 dr}{\sqrt{1 - r^2/R^2 + A_2/r}} \tag{41}$$

The integral of (41) is difficult. If the third item is neglected, we obtain [1]:

$$V' \approx 4\pi \int_{R_1}^{R_2} \frac{r^2 dr}{\sqrt{1 - r^2/R^2}} = \frac{4\pi R^3}{3}\left(\arcsin\frac{r}{R} - \frac{r}{R}\sqrt{1 - \frac{r^2}{R^2}}\right)\Bigg|_{R_1}^{R_2}$$

$$= \frac{4\pi}{3}\left(R_2^3 - R_1^3\right) + \frac{2\pi}{5R^2}\left(R_2^5 - R_1^5\right) + Q(R_2, R_1, R) \tag{42}$$

In this case, let's estimate the magnitude of volume's change in curved space. Let $V' = V_0 + \Delta V$ and $R_1 = 0$ in (42), by omitting high order items, we have

$$\frac{\Delta V}{V_0} = \frac{3R_2^2}{10R^2} = \frac{4\pi G \rho_0 R_2^2}{5c^2} \tag{43}$$

For neutron stars, we have $\rho_0 \sim 10^{15}$ and $R_2 \sim 10^4$, so $\Delta V/V_0 = 1.8 \times 10^{-4}$. If we consider the universe as a uniform sphere, we have $\rho_0 \sim 10^{-27}$ and $R_2 \sim 10^{26}$, so $\Delta V/V_0 = 1.8 \times 10^{-2}$. For so-called black hole, according to (4), we have $R_2 = c^2/(3\pi G \rho_0)$ and $\Delta V/V_0 = 4/15 = 0.27$. For common spheres, $\rho_0 \Delta V$ is a very small but non-zero quantity. Therefore, in the curved space, we have

$$M \neq \frac{4\pi G \rho_0}{3}\left(R_2^3 - R_1^3\right) \tag{44}$$

On the other hand, in order to obtain the relation among gravity mass, radius and density, we need to discuss the definition of gravity mass. According to general relativity, the energy of static gravity field is calculated by following formula[1]:

$$E = cP_0 = \int \sqrt{-g}\left(T_0^0 - T_1^1 - T_2^2 - T_3^3\right)dx_1 dx_2 dx_3 \tag{45}$$

It has been proved in general relativity that when $A_2 = 0$, by using the metric (23), we can obtain [1]

$$\rho_g(r) = \rho_0\left(1 - \frac{r^2}{R^2}\right) \qquad M_g = \frac{4\pi G \rho_0}{3}\left(R_2^3 - R_1^3\right) \tag{46}$$

Here $\rho_g$ is the density of gravity mass. In this case, the gravity mass of general relativity in curved space is the same as that of the Newtonian theory in flat space. However, because we have $A_2 \neq 0$ for hollow sphere, the formula (46) does not hold. Let's discuss this problem in detail. By using metric (32), we have

$$\sqrt{-g} = \frac{F(r)r^2 \sin\theta (C_2 + B_2 D(r))}{\sqrt{r}\sqrt{1 - r^2/R^2 + A_2/r}} \tag{47}$$

By considering (9) and relation $dV' = dV/\sqrt{1 - r^2/R^2 + A_2/r}$ in curved space, (45) can be written as

$$E = \int \frac{F(r)(C_2 + B_2 D(r))(\rho_0 c^2 + 3p(r))}{\sqrt{r}} dV' \tag{48}$$

By using (16) and (31), the density of gravity mass in general relativity can be written as

$$\rho_g(r) = \frac{F(r)(C_2 + B_2 D(r))(\rho_0 c^2 + 3p(r))}{c^2 \sqrt{r}} = \rho_0\left[1 - \frac{\sqrt{r} + 2F(r)(C_2 + B_2 D(r))}{\sqrt{r}}\right] + \frac{3B_2}{c^2} \tag{49}$$

So the gravity mass of hollow sphere should be

$$M_g = \int \rho_g(r) dV' = \int_{R_1}^{R_2} \frac{4\pi \rho_g(r) r^2 dr}{\sqrt{1 - r^2/R^2 + A_2/r}} \tag{50}$$

If the item $A_2/r$ is neglected, we obtain

$$M_g = \frac{4\pi \rho_0}{3}\left(R_2^3 - R_1^3\right) + \frac{2\pi \rho_0}{5R^2}\left(R_2^5 - R_1^5\right) + Q(R_2, R_1, R)$$

$$+ \int\left[\frac{3B_2}{c^2} - \rho_0 \frac{\sqrt{r} + 2F(r)(C_2 + B_2 D(r))}{\sqrt{r}}\right]dV' \tag{51}$$

Because the item $A_2/r$ exists actually which has the same magnitude as the item $r^2/R^2$, the form of $M_g$ becomes very complex. However, we can always write it as

$$M_g = \frac{4\pi \rho_0}{3}\left(R_2^3 - R_1^3\right) + \Delta M \tag{52}$$

We may have $\Delta M > 0$ or $\Delta M < 0$. On the other hand, we do not distinguish gravity mass and inertial mass in the Newtonian mechanics, so $M$ in (7) can be both gravity mass and inertial mass. In the Einstein's theory of gravity, gravity mass and inertial mass are connected by the principle of equivalence. According to the principle of equivalence, gravity mass is equivalent to inertial mass. The inertial mass here is just that defined in the Newtonian mechanics. So we have $M_g = M$. That is to say, if the gravity mass of a static hollow sphere with energy momentum tensor (9) defined in general relativity is $M_g$, the mass is just equal to the gravity mass of the Newtonian mechanics. Substitute (52) into (39), we have

$$A_1 = -\frac{2G\Delta M}{c^2} \tag{53}$$

Because $R_1$, $R_2$ and $\rho_0$ can be chosen arbitrarily, we have $\Delta M \neq 0$ and $A_1 \neq 0$ in generally. Using (53) in (37) and considering (36), we obtain

$$B_2 = \frac{\sqrt{R_2 - 2GM/c^2}/F(R_2) - \sqrt{R_1 - 2G\Delta M/c^2}/F(R_1)}{D(R_2) - D(R_1)} \tag{54}$$

$$C_2 = \frac{D(R_1)\sqrt{R_2 - 2GM/c^2}/F(R_2) - D(R_2)\sqrt{R_1 - 2G\Delta M/c^2}/F(R_1)}{D(R_1) - D(R_2)} \tag{55}$$

Similarly, because $R_1$, $R_2$ and $\rho_0$ are arbitrary, we have $B_2 \neq 0$ and $C_2 \neq 0$ in general. In this way all integral constants are determined. Substitute (52) into (33), in the vacuum region $I_1$ of spherical cavity, we have

$$g_{00}(r) = \left(1 - \frac{2G\Delta M}{c^2 r}\right) \qquad g_{11}(r) = -\left(1 - \frac{2G\Delta M}{c^2 r}\right)^{-1} \tag{56}$$

So there are singularities at the spherical center $r = 0$. The result is different from that of the Newtonian theory. According to the Newtonian theory, the material distributed outside the spherical cavity with spherical symmetry does not affect the gravity field in the cavity. But according to (56), it does in curved space.

It should be emphasized that according to the theory of differential equation, we should determine the integral constants by the initial and boundary conditions which are known. For the problem of spherical symmetry, only the gravity field which is far from the center of sphere is known. The metric at the center of sphere is unknown until the integral constants are determined. It is unsuitable for us to take $A_1 = 0$ because it is not boundary condition. The result would destroy the continuity of metric on the boundary and lead to the insistence of theory. In fact, if let $A_1 = 0$ in (33), on the internal spherical surface with $r = R_1$, we have

$$\frac{F^2(R_1)}{R_1}\left[C_2 + B_2 D(R_1)\right]^2 = 1 \qquad \left(1 - \frac{R_1^2}{R^2} + \frac{A_2}{R_1}\right) = 1 \tag{57}$$

So we still have $A_2 = R_1^3/R^2 \neq 0$. Therefore, the formulas (41) ~ (51) hold and the gravity mass is still described by (51). On the external spherical surface $r = R_2$, the continuity condition becomes

$$\frac{F^2(R_2)}{R_2}\left[C_2 + B_2 D(R_2)\right]^2 = 1 - \frac{2GM}{c^2 R_2} \qquad \left(1 - \frac{R_2^2}{R^2} + \frac{R_1^3}{R^2 R_2}\right) = 1 - \frac{2GM}{c^2 R_2} \tag{58}$$

Form (58) we have

$$M = \frac{4\pi G \rho_0}{3}\left(R_2^3 - R_1^3\right) \tag{59}$$

However, we have proved that (59) does not hold. In curved space, according to (51), we should have

$$M_g = M \neq \frac{4\pi G \rho_0}{3}\left(R_2^3 - R_1^3\right) \tag{60}$$

It is obvious that $A_1 = 0$ violates the boundary continuity and theoretical consistence.

# 4. The singularity of the inner metric of hollow sphere

According to (56), the metric has singularity at the point $r = 0$. This is inherent singularity which can not be eliminated by the coordinate transformation. The seriousness of the problem is that for any hollow sphere composed of common material, no matter what its mass and density are, a singularity exists in its center. This result does not coincide with practical observation and is completely impossible.

On the other hand, we consider (14) and let

$$1 - \frac{r^2}{R^2} + \frac{A_2}{r} = 0 \qquad \text{or} \qquad r^3 - R^2 r - R^2 A_2 = 0 \qquad (61)$$

It seems that there is a singularity surface beneath two spherical shells. We now discuss this problem. The real solution of (61) is

$$r = \left[ \frac{R^2 A_2}{2} + \sqrt{\left(\frac{R^2 A_2}{2}\right)^2 - \left(\frac{R^2}{3}\right)^3} \right]^{\frac{1}{3}} + \left[ \frac{R^2 A_2}{2} - \sqrt{\left(\frac{R^2 A_2}{2}\right)^2 - \left(\frac{R^2}{3}\right)^3} \right]^{\frac{1}{3}} \qquad (62)$$

However, if $r$ is really a real number, the following relation should be satisfied

$$\delta = \left(\frac{R^2 A_2}{2}\right)^2 - \left(\frac{R^2}{3}\right)^3 \geq 0 \qquad (63)$$

According to (35) and (51), we have

$$R^2 A_2 = -\frac{2GMR^2}{c^2} + R_2^3 = R_1^3 - \frac{2G\Delta MR^2}{c^2} \qquad (64)$$

In the weak field to neglect the item $2G\Delta MR^2/c^2$, we can obtain

$$R_1 \geq \left(\frac{2}{\sqrt{27}}\right)^{\frac{1}{3}} \left(\frac{3c^2}{8\pi G \rho_0}\right)^{\frac{1}{2}} = \frac{9.26 \times 10^{12}}{\sqrt{\rho_0}} \qquad (65)$$

We know that even for high-density celestial body just as white dwarf, the difference is still small to calculate based on general relativity and the Newtonian theory. The material density of white dwarf is $\rho_0 = 10^{11} Kg/m^3$. According to (65), we have $R_1 \geq 2.7 \times 10^7 m$ which has the same magnitude with the radius of white dwarf. For common galaxy, we have $\rho_0 = 10^{-21} Kg/m^3$. According to (65), we have $R_1 \geq 3 \times 10^{22} m$ which is just the size of galaxies. So (64) can also be satisfied. In this way, (62) becomes

$$r = \left( \frac{R_1^3}{2} - \frac{G\Delta M}{c^2} + \sqrt{\delta} \right)^{\frac{1}{3}} + \left( \frac{R_1^3}{2} - \frac{2G\Delta M}{c^2} - \sqrt{\delta} \right)^{\frac{1}{3}} \qquad (66)$$

By developing (66) into the Taylor series, it can see that if $\Delta M > 0$, we have $r < R_1$. That is to say, the singularity surface is in the cavity. Because the metric of cavity is (33), in stead of (14), there is no singularity surface in the region $R_1 < r < R_2$ inside the hollow sphere. If $\Delta M < 0$, we have $r > R_1$. There is no singularity surface in the region $R_1 < r < R_2$ inside the hollow sphere. Besides, there is also a singularity surface outside the external surface of hollow sphere according to (8), which is so-called horizon. We do not discuss it in this paper. According to (16), the pressure intensity in the hollow sphere is

$$p(r) = B_2 e^{-\nu(r)/2} - \rho_0 c^2 = \frac{B_2 \sqrt{r}}{F(r)(C_2 + B_2 D(r))} - \rho_0 c^2 \qquad (67)$$

If there is a surface inside the hollow sphere with radius $r = r'$ on which we have

$$C_2 + B_2 D(r') = 0 \quad (68)$$

the pressure intensity on the surface would become infinite. Therefore, if there is black hole in hollow sphere, the black hole could only be a spherical surface. Substitute (53) and (54) in (68), we get

$$D(r') = \frac{D(R_2)\sqrt{R_1 - 2G\rho_0 \Delta V/c^2}/F(R_1) - D(R_1)\sqrt{R_2 - 2GM/c^2}/F(R_2)}{\sqrt{R_1 - 2G\rho_0 \Delta V/c^2}/F(R_1) - \sqrt{R_2 - 2GM/c^2}/F(R_2)} \quad (69)$$

Because the radii $R_1$ and $R_2$ of hollow sphere are arbitrary, it can be seen from the form of (69) that we may find a proper $r'$ so that (68) can be satisfied. However, on this spherical surface composed of black holes, space-time has no singularity. That is to say, the surface of space-time singularity does not overlap with the surface on which material collapses. This is incomprehensible. It should be noted that up to now we have no any restriction on the mass and density of hollow sphere. This result means that even for common hollow spheres, they may be unstable. They may collapses into spherical surface's black holes!

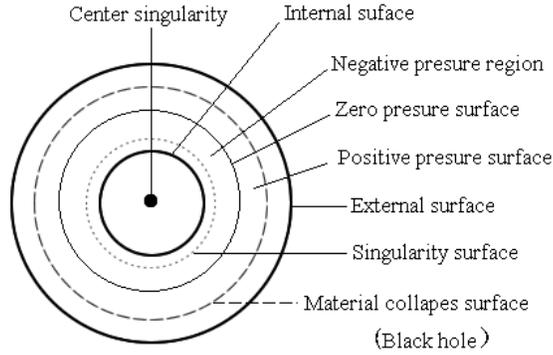

**Fig.1 The singularity of hollow sphere**

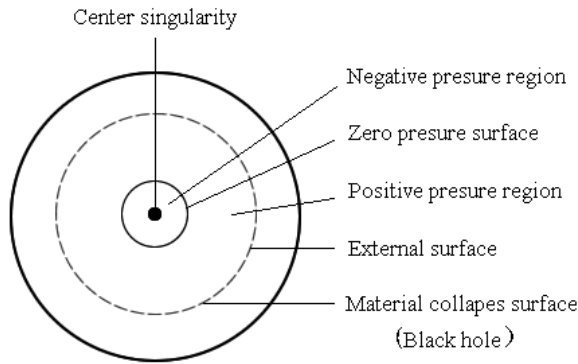

**Fig.2 The singularity of solid sphere**

Similarly, because the internal and external radii are arbitrary, let $r = R_2$ or $r = R_1$, we have $p(R_2) \neq 0$ and $p(R_1) \neq 0$ in general. Because the hollow sphere is placed in vacuum without material outside and inside its two surfaces, this result is also incomprehensible. The singularities of hollow sphere are shown in fig.1. It is obvious that all of them can not be true.

# 5. The singularities of solid sphere's metric and black holes

According to the present calculation of general relativity, the internal metric of a common solid sphere has no singularity when the radius of sphere is greater than the Schwarzschild radius. According to the strict calculation in this paper, the situation is completely different. The solid sphere is a special situation of hollow sphere when its internal radius becomes zero. If the internal radius of hollow sphere is not considered, we obtain the results of solid sphere. The internal metric of solid sphere is still described by (30), but the conditions of boundary are different. On the spherical surface $r = R_2$, we have

$$1 - \frac{2GM}{c^2 R_2} = 1 - \frac{R_2^2}{R^2} + \frac{A_2}{R_2} \tag{70}$$

$$1 - \frac{2GM}{c^2 R_2} = \frac{F^2(R_2)}{R_2}\left[C_2 + B_2 D(R_2)\right]^2 \tag{71}$$

From (70) we can determine $A_2$, and the result is the same as (35). Similarly, because the volume of sphere in curved space is different from that in flat space, according to (41) with $R_1 = 0$, we have

$$V \neq \frac{4\pi G}{3} R_2^3 \tag{72}$$

Therefore, we also have $A_2 \neq 0$. In order to obtain the relation among mass, density and radius of a sphere in curved space, we consider the definition of gravity mass. By the same procedure to obtain (45) ~ (51), we also have

$$M_g = M = \frac{4\pi \rho_0}{3} R_2^3 + \Delta M \tag{73}$$

Substitute it in (70), we get $A_2 = -2G\Delta M R^2/c^2$. However, we can not determinate $C_2$ and $B_2$ simultaneously only based on (71). Another condition is needed. By considering the fact that the pressure intensity on the surface of sphere should be zero with $p(R_2) = 0$, we have from (16)

$$\sqrt{R_2} B_2 = \rho_0 c^2 F(R_2)\left[C_2 + B_2 D(R_2)\right] \tag{74}$$

From (71) and (74), we have

$$B_2 = \rho_0 c^2 \sqrt{1 - \frac{2GM}{c^2 R_2}} \qquad C_2 = \sqrt{1 - \frac{2GM}{c^2 R_2}}\left[\frac{\sqrt{R_2}}{F(R_2)} - \rho_0 c^2 D(R_2)\right] \tag{75}$$

Now, all integral constants are determined. The internal metric of solid sphere is

$$ds^2 = \frac{F^2(r)}{r}\left[C_2 + B_2 D(r)\right]^2 dt^2 - \left(1 - \frac{r^2}{R^2} + \frac{A_2}{r}\right)^{-1} dr^2 - r^2\left(d\theta^2 + \sin^2\theta\, d\varphi^2\right) \tag{76}$$

It is obvious that $g_{00}(0)$ has singularity at $r = 0$. According to (30), we have $F(0) \neq 0$. So we also have $C_2 + B_2 D(0) \neq 0$, unless

$$\rho_0 c^2 \left[D(R_2) - D(0)\right] = \frac{\sqrt{R_2}}{F(R_2)} \tag{77}$$

However, because $R_2$ is arbitrary, (77) can not hold in general situation. So $g_{00}(0)$ is infinite at the center of sphere. In order to avoid infinite, according to the current calculation in general relativity, we directly let $A_2 = 0$ in (14) and obtain the metric (23). Then obeying the calculation procedure of (45) ~ (51) and get

$$M_g = M = \frac{4\pi G \rho_0}{3}\left(R_2^3 - R_1^3\right) \tag{78}$$

In this way, the infinite at the center of sphere seems to be avoided. However, as mentioned before, several serious problems will be caused. We discuss this problem further below.

1. According to the theory of differential equation, the integral constant should be determined by the known initial and boundary conditions. For the gravity field with spherical symmetry, what can be certain is the gravity potential which satisfies the Newtonian formula at the place far from the center of sphere. By considering the continuity of metric, the integral constant can only be determined by the boundary condition on the spherical surface. We can not take the metric of spherical center as the boundary condition, because the metric at the spherical center is unknown before the integral constant has not been determined. We have no any transcendental reason to let $A_2 = 0$, for the theory itself may cause singularity at the center of sphere. In fact, according to the theme of singularity proved by S. W. Hawking etc., the singularity can not be avoided in the Einstein's theory of gravity. The theme has no any restriction for the mass and density of a body. For hollow and solid spheres composed of common material, singularities may exist according the Einstein's theory of gravity! We will discuss the theme of singularity in next section further.

2. The uniqueness of theory will be violated if we to let $A_2 = 0$ directly. It means that we can use two methods to determine integral constant $A_2$. One is to use the continuity condition on the spherical surface and another is to use the limitation condition of metric at the center of sphere. But these two methods would cause two different results so that it can not be allowed in physics.

3. It would cause the inconsistence of hollow and solid sphere's metrics if we let $A_2 = 0$ directly. In fact, if let $A_2 = 0$ directly according to general relativity, we have $g_{00} = -g_{11}(1) = 1$. It means that the space at spherical center is flat. However, it is impossible for the metric of hollow sphere to let $A_2 = 0$. According to (32), the space-time curvature is infinite at the center of hollow sphere. However, the solid sphere is only the special situation of hollow sphere with a zero inner radius. It is impossible for them to have such great difference. In fact, we can make the vacuum cavity of hollow sphere small enough, so that the hollow sphere can be considered as a solid sphere actually. For example, there is very small vacuum bubble at the center of a cast iron sphere. Do we consider the cast iron as hollow or solid spheres? For the cast iron sphere without vacuum bubble, the curvature at its center is zero. How can the curvature become infinite when a very small vacuum bubble appears at its center?

Let's discuss whether or not there is other singularity inside solid sphere. According to (70) and (73), we have

$$R^2 A_2 = -\frac{2GMR^2}{c^2} + R_2^3 = -\frac{2G\Delta MR^2}{c^2} = \frac{3\Delta M}{2\pi\rho_0} \tag{79}$$

According to (62) and (63), if $r$ is a real number, we should have

$$\delta = \left(\frac{3\Delta M}{2\pi\rho_0}\right)^2 - \left(\frac{c^2}{8\pi G \rho_0}\right)^3 \geq 0 \tag{80}$$

By simple calculation to let $V' = V + \Delta V$ and $\Delta M = \rho_0 \Delta V$, based on (42), we have $\Delta V = 2\pi R_2^5/(5R^2)$

as well as

$$\sqrt{\rho_0} \geq \frac{5^{1/5}c}{\sqrt{8\pi G}R_2} = \frac{2.33\times 10^{13}}{R_2} \qquad (81)$$

For common objects, (81) can not be satisfied. To take $R_2 = 10^5 m$, we get $\rho_0 = 5\times 10^{16} Kg/m^3$ based on (81). The density is similar to that of neutron star. For such objects, we have

$$R^2 = \frac{3c^2}{8\pi G\rho_0} = 3.32\times 10^9 \qquad R^2 A_2 = \frac{3\Delta V}{2\pi} = \frac{8\pi G R_2^5 \rho_0}{5c^2} = 1.86\times 10^{15} \qquad (82)$$

$$\delta = \left(\frac{R^2 A_2}{2}\right)^2 - \left(\frac{R^2}{3}\right)^3 = 8.65\times 10^{29} - 1.22\times 10^{28} = 8.64\times 10^{29} \qquad (83)$$

According to (62), we obtain

$$r = \left(\frac{R^2 A_2}{2} + \sqrt{\delta}\right)^{\frac{1}{3}} + \left(\frac{R^2 A_2}{2} - \sqrt{\delta}\right)^{\frac{1}{3}} \approx 2\left(\frac{R^2 A_2}{2}\right)^{\frac{1}{3}} = 2.46\times 10^5 m > 10^5 \qquad (84)$$

Therefore, besides the singularity appearing at the center of sphere, it seems that no other singularity exists inside a stable solid sphere.

For a solid sphere, the internal pressure intensity is also described by (67), but integral constant should be described by (75). We have $F(0) \neq 0$ and $C_2 + B_2 D(0) \neq 0$ in general. So according to (67), we have $p(0) = -\rho_0$. That is to say, the pressure intensity at the center of sphere can not be infinite. No matter what are the mass and density of sphere, the material does not collapse toward the center of sphere. In stead, the pressure intensity becomes negative value. Besides, in the region nearing the center of sphere, the pressure intensity is also a negative value. On the other hand, if there is spherical surface with radius $r = r'$, on which we have

$$C_2 + B_2 D(r') = 0 \qquad (85)$$

The pressure intensity would become infinite on it. Substitute (75) in(85), we have

$$D(r') = D(R_2) - \frac{\sqrt{R_2}}{\rho_0 c^2 F(R_2)} \qquad (86)$$

It is seen form (86) that such surface may exist. Therefore, if black hole exists in solid sphere, the black hole can only be a spherical surface. Such solid sphere is not stable for material would collapse toward into a spherical surface. However, there is no singularity of space-time curvature on the spherical surface. Such result is also uncanny.

The singularity of solid sphere is shown in Fig. 2. It is noted that though we have no any restriction for mass and density, so many strange characters appear for a common solid sphere. The results are completely different from the current conclusions of general relativity. Because we can not let integral constant $A_2 \neq 0$ in the metric of hollow and solid spheres, the theory of singularity black hole in the current astrophysics and cosmology is overthrown thoroughly.

# 6．Discussions on the theme of singularity and the rationality of singularity black hole theory

Hawking etc. proved the theme of singularity by means of the method of differential geometry. The theme was based on three prerequisite conditions[2]. 1. General relativity was tenable. 2. The law of causality was tenable. 3. There were some points in space-time at which material density was non-zero. The theme declaimed that if theses three conditions were satisfied, singularity inevitably existed in space-time. Hawking and Penrose considered singularities as the beginning and ending points of time which endow a new meaning for singularity. The Big Bang theory was considered as the beginning of time and the black holes were regarded as the ending of time. It notes that the theme has no restriction on material's mass and density and not demands that singularities are embodied in material. That is to say, according to the theme, singularities may be bared in vacuum. In order to avoid this embarrassing situation, Penrose proposed the so-called principle of the universe supervisor. The principle declared that there exist the universe supervisors who prohibit the appearance of bare singularities in vacuum. In other word, due to the existence of the universe supervisors, all singularities would be wrapped in the centers of black holes with great masses and high densities. According to the solutions of the Einstein's equation of gravity, there exist the Schwarzschild black holes with spherical symmetry and the Kerr black holes with axial symmetry and so on. Theses black holes with singularities were hidden in the centers of material with big masses and high densities. Because they can not be perceived directly, physicists may tolerate their existence.

However, the strict calculation in this paper reveals that the universe supervisors can neither avoid the appearance of singularities in vacuum, nor avoid their appearance at the centers of common hollow and solid spheres with small masses and low densities. In order to avoid the singularity appearing in the center of sphere, the current calculation always let the integral constant to be zero directly. According to the accurate calculation in this paper, the integral constant can not be zero. We should determine the integral constant by the metric continuity on the boundaries, in stead of supposing them to be zero. By considering the fact that the volume of sphere in curved space is different from that in flat space, the integral constant can not be zero. Therefore, no matter what the mass and density of a hollow and solid sphere are, singularity would appear inevitably at their centers according to general relativity. On the other hand, because the pressure intensity can not be infinite at the center of sphere, material can not collapse towards the spherical center. In stead, the pressure intensity would become negative values at the center and the nearby region. Meanwhile, there may have a curved surface inside the hollow and solid sphere on which pressure intensity is infinite so that material would collapse toward it. But the space-time curvature is still finite on the surface. It means that space-time singularity surface dose not overlap with the curved surface on which the pressure intensity is infinite. Meanwhile, for the hollow spheres composed of common material, the pressure intensities on the external and internal surfaces would not be zero. All these characters can not coincide with practices and incomprehensible in physics.

In the paper "Singularities of the Gravitational Fields of Static Thin Loop and Double Spheres", the author further proved that by means of the coordinate transformations of the Kerr solution and the Kerr-Newman solution of the Einstein's equation of gravitation field with axial symmetry, the gravitational fields of the static thin loop and double spheres are obtained. The results indicate that, no matter what mass and density are, there are singularities at the central point of thin loop and the contact point of two spheres. What is more, the singularities are completely exposed in vacuum. Space near the surfaces of thin loop and spheres are highly curved, even though the gravitational fields are very weak. These results are consistent with the calculation of hollow sphere in this paper in which no any material exists in cavity the so that the singularity is exposed in vacuum.

In fact, the three conditions of the theme of singularity do not contain the restrictions of mass and density of material systems without demanding large mass and high density. That is to say, according to the theme singularity, it is certain to have space-time singularity points for common objects in daily life just as hollow spheres, solid spheres and thin rings composed of common material with small masses and low densities. But common knowledge tells us that this is untrue. Any practical experiment and measurement can deny this result. Traditional theory considers that space-time singularity and black holes appear in the centers of celestial bodies with great mass and high density. It is incomprehensible for singularities to appear in material system with small mass and low density. So it is certain that at least one of three conditions is untenable. The existence of material and the law of causality are the foundation of science and our world. So only possibility is that general relativity has something wrong. All singularity black holes, white holes and worm holes in the current cosmology and astrophysics have nothing to do with real world and not exist in nature. The singularities appearing in general relativity are caused by the descriptive method of mathematics actually, instead of causing by large masses and high densities of material. The black holes are considered to exist in the center of Quasars according to the current theory. However, according to the observations of Rudolf E. Schild and Darryl J. Leiter (Rudolf E. Schild, Darryl J. Leiter, 2006), the centre of Quasar 0957+561 is a close object, called a MECO (Massive Eternally Collapsing Object). Not a singularity black hole, it is surrounded by a strong magnetic field. The observation of Rudolf E. Schild was consistent with the calculation and analyses in this paper. If there sre black holes in the universal space, they can only be the Newtonian black holes, not the Einstein's singularity black holes!

More essentially, the true world excludes infinites. A correct theory of physics can not tolerate the existence of infinites. It is well known that the history of physics is the one to overcome infinites. Modern physics grows up in the process to surmount infinites. As revealed in this paper, singularity in general relativity is actually caused by the description method of curved space-time. Physicists and cosmologists should take cautious and incredulous attitude on the problems of black holes. It is not a scientific attitude to consider singular black holes as objective existence without any question on them. We should think in deep, whether or not our theory has something wrong. When we enjoy the beauty and symmetry of the Einstein's theory of gravity, remember that we should not neglect its limitations and possible mistake.